# Joint discovery of haplotype blocks and complex trait associations from SNP sequences

Nebojsa Jojic
Microsoft Research
Redmond, WA 98052

Vladimir Jojic
Microsoft Research
Redmond, WA 98052

David Heckerman
Microsoft Research
Redmond, WA 98052

## Abstract

Haplotypes, the global patterns of DNA sequence variation, have important implications for identifying complex traits. Recently, blocks of limited haplotype diversity have been discovered in human chromosomes, intensifying the research on modelling the block structure as well as the transitions or co-occurrence of the alleles in these blocks as a way to compress the variability and infer the associations more robustly. The haplotype block structure analysis is typically complicated by the fact that the phase information for each SNP is missing, i.e., the observed allele pairs are not given in a consistent order across the sequence. The techniques for circumventing this require additional information, such as family data, or a more complex sequencing procedure. In this paper we present a hierarchical statistical model and the associated learning and inference algorithms that simultaneously deal with the allele ambiguity per locus, missing data, block estimation, and the complex trait association. While the block structure may differ from the structures inferred by other methods, which use the pedigree information or previously known alleles, the parameters we estimate, including the learned block structure and the estimated block transitions per locus, define a good model of variability in the set. The method is completely data-driven and can detect Chron's disease from the SNP data taken from the human chromosome 5q31 with the detection rate of 80% and a small error variance.

## 1 Introduction

Recently, blocks of limited haplotype diversity have been discovered in human chromosomes [1, 2, 3]. These blocks consist of a number of polymorphic nucleotide sites separated from each other by nucleotides which do not vary in the human population. For a given haplotype block only a small subset of configurations are observed compared to the whole space of possibilities $4^{length\ of\ block}$. Several hypotheses on processes that gave rise to such block structure have been described in the literature: presence of recombination hotspots[19], founder effects/bottleneck [10] and genetic drift[18]. Differences in frequency of particular states of haplotype blocks have been associated with susceptibility to different diseases [2, 20]. These observations led to intensified interest in modelling the block structure as well as the transitions or co-occurrence of the alleles in these blocks as a way to compress the variability and infer the associations more robustly [4, 6, 7, 8, 9, 13, 12].

The haplotype block structure analysis is typically complicated by the fact that the phase information for each SNP is missing, i.e., the observed allele pairs are not given in a consistent order across the sequence. The techniques for circumventing this require additional information, such as family data, or a more complex sequencing procedure [1].

In this paper, we present a hierarchical statistical model and the associated learning and inference algorithms that simultaneously deal with the allele ambiguity per locus, missing data, block estimation, and the complex trait association. Given only a set of sequences of unordered pairs of nucleotides, and *without the family data*, we are able to learn appropriate blocks of alleles and estimate the missing data by deriving a variant of the EM algorithm [16] for our model. While the block structure may differ from the structures inferred by other methods, which use the pedigree information or previously known alleles, the parameters we estimate, including the learned block structure and the estimated block transitions per locus, define a good model of variability in the set. This is illustrated by the classification rates: After training one model on SNPs from people that contracted the disease and another on the SNPs from the healthy individuals, we can use the models to infer the probability of contracting the disease given a new sequenced genotype. The method can detect Chron's disease from the SNP data taken from the human chromosome 5q31 with the detection rate of 80% and a small error variance.

## 2 SNP sequence modelling

Ultimately, we aim to model the distribution over the possible configurations of SNPs and their joint probability distribution with various complex traits. In theory, by collecting enough statistical data this should be possible. However, in practice, the data is scarce and the computational complexity of the brute force approach is immense. Thus, incorporating the biological knowledge into the distribution modelling is necessary to reduce the requirements both on computational complexity and the size of the dataset.

Our approach to such modelling is based on graphical models, hierarchical statistical models that describe the target distribution with the aid of a variety of hidden variables, each meant to capture a certain aspect of the variability in the data. The variables are connected into a network of conditional dependencies that reflects the known structure of the data. One good way to build graphical models is to develop a simplified generative process that can generate the data in accordance with the knowledge about the domain. The generative model captures the probability distribution over the variables of interest through parameters of the conditional probability distributions, which can be learned by maximizing the likelihood of generating the data from a training set.

Once the generative model is fitted to the training set, it can be used both on the previously seen and the unseen (new) data to separate causes of variability associated with the hidden variables and model parameters.

The interesting causes of variability in SNP sequences are:

- A small number of characteristic patterns $R_s$, $s = 1, ..., S$ that can be used in a block structure to represent a large fraction of observed sequences

- The labels assigning observed SNPs to the patterns

- The missing data

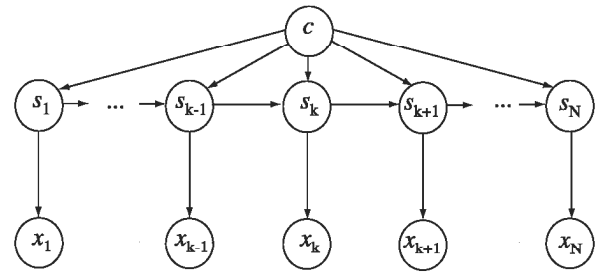

Figure 1: A generative model that allows learning the block structure of haplotypes, while also providing the framework for complex trait prediction

- The uncertainty in the nucleotide pair assignment to the chromosome pair

- Other sources of noise causing incorrect measurements

- The complex traits associated with the region being studied

In the generative model in Fig. 1, the complex traits of interest are indexed by $c$ (in our experiments, for example, $c = 1$ indicates the presence of the Chron's disease and $c = 0$ indicates a healthy individual). This variable has a prior distribution $p(c)$ defined by the expected probability distribution of the trait(s) in the population. This parameter can also be estimated together with other model parameters from the data, but we simply set it to uniform ($p(c = 1) = p(c = 0) = 0.5$). The variable $c$ will govern the distribution over the other variables in the system, by affecting the distribution over the combinations of the haplotype blocks. This distribution is governed by a Markov chain of the pattern index variables $s_k$, associated with the observed nucleotide $x_k$. In the hidden Markov chain, the transition probabilities $p_k(s_k|s_{k-1}, c) = \theta_{kc}^{s_k s_{k-1}}$ are affected by the choice of class variable $c$. We assume for the moment that we observe a single nucleotide $x_k$, whose distribution is given as $p(x_k|s_k) = p(x_k|R_k^{s_k})$, i.e., the observation at the k-th locus depends on the value of the $k - th$ element in the $s_k$-th pattern in the library (see Table 1 for illustration). This dependence is generally noisy for two reasons: to model all the remaining variability, and to provide for slower convergence in early iterations of the EM algorithm, thus avoiding local maxima. The noise parameter can be annealed or estimated from the data, as the expected number of times the observed nucleotide did not match the expected pattern at the locus. The patterns are assumed to be the same length as the sequences. However, there is typically many fewer patterns than the observed sequences, making the library learning possible. For in-stance, in our experiments, we use a small library of 7 patterns to model 764 sequences. From previously published work, we inferred that the data typically had a diversity of up to 4 different haplotype blocks at any site, and so we used a larger number to avoid local maxima.

One can think of every given SNP sequence as having been generated by sampling from this generative model:

- Draw the trait class $c$ from the distribution $p(c)$

- Based on the class $c$, draw the first pattern index variable $s_1$ according to $\theta_{1c}$ and then generate a nucleotide $x_k$ by drawing from $p(x_1|R_1^{s_1})$

- Based on $s_1$ and c, draw the second pattern index variable $s_2$ according to $\theta_{1c}^{s_1 s_2}$ and generate a nucleotide $x_k$ by drawing from $p(x_2|R_2^{s_2})$

- Continue down the chain until all variables are generated

Note that the above parametrization does not necessarily force block structure in the generated data. However using appropriate transition matrices, those that allow transitions between patterns at particular positions in a sequence (recombination hotspots), such block structure is generated.

If we define the pattern library using the prior knowledge of the alleles, the above method would generate haplotypes in a block structure. Since we estimate the patterns from the data together with the segmentation $s_k$ and other hidden variables and model parameters, and since we do not use the family data, the learned patterns may not correspond to optimal haplotype blocks, although in our experiments, they do provide a block structure for the data similar to the one reported in previous work. The rationale behind such analysis is that incorporating the block-structure assumption proposed by the human genome researchers into the model, learning the joint probability distribution over the traits and SNP sequences will be possible on smaller datasets and using tractable algorithms. With the goal of capturing the distribution over the variable, rather than finding the optimal block segmentation, the learning algorithm, as we will discuss, can find many local maxima with slight variations in likelihood that provide different sequence segmentations, but with similar trait classification/detection rates, as many sets of patterns and transition probabilities are equivalent in terms of the distribution they define.

For example, in Table 1, we show a short pattern library, the important transition probabilities and the sequences that the model can generate according to these parameters. In most of the sequence, the transition probabilities favor staying in the same pattern, and only at locus 3, a jump is possible. Next to each pattern, we show the probability of generating each of the possible four sequences. A variety of qualitatively similar pattern libraries can be defined so as to yield the same distribution. A trivial example is the exchange of the 4-letter end blocks between two patterns with the appropriate change in the transition probabilities. Thus, none of the patterns in the library has to be a valid haplotype. To study the haplotype block diversity, it is necessary to compute the regions of label constancy (transition matrices favoring staying in the same patters) and then identify the possible blocks form the pattern library that can occupy these positions in the sequences. The transitions among blocks then complete the standard picture of the haplotype diversity under the block-structure assumption.

Another, perhaps less intuitive way of perturbing the library and transition probabilities without changing the pdf is through letter or segment exchange within a block. For example, at the second locus, we can exchange letters C and A in the first and the second pattern and then set the transition probabilities for sites 1 and 2 to be: $\theta_1^{12} = 1, \theta_2^{21} = 1, \theta_1^{21} = 1, \theta_2^{12} = 1$. If such a model is learned from the data, the switch can again be detected in the post-processing step if the model is to be used to identify long contiguous haplotype blocks. However, from the modelling point of view, this is usually not necessary

This ambiguity in parameters is helpful as it helps avoid local maxima in estimating an appropriate pdf through these parameters. On the other hand the overparameterization can lead to overtraining. Previous research on haplotype block structures allows us to initialize the transition costs to favor staying within the same pattern, thus avoiding some of these problems, as we will discuss later.

### 2.1 Inference and learning

Typically, in a graphical model, the inference process is defined as computing the posterior distribution over the hidden variables (in our case $\{s_k\}_{k=1}^N$ and c) given the observed data $\{x_k\}_{k=1}^N$, i.e., $p(c, \{s_k\}_{k=1}^N | \{x_k\}_{k=1}^N)$ In particular, computing the marginal $p(c = 1 | \{x_k\}_{k=1}^N)$ provides the estimate on the probability of contracting the disease given the SNP sequence of interest. This marginal involves integrating over all the uncertainty involved in breaking the data into blocks from the learned pattern library.

To compute the posterior, we observe that for a fixed c, the model has the form of a hidden Markov model, and thus $p(\{s_k\}_{k=1}^N | \{x_k\}_{k=1}^N, c)$ can be efficiently rep-

Table 1: An example of using a generative model in Fig. 1 to describe a (single-class) distribution over possible SNP sequences. For given parameters, the model can generate only 4 different sequences. Two blocks are evident in each state sequence, comprising of the first three sites and the last four. A different pattern library can easily be constructed that reveals the same block structure but stores the prototypes in different parts of the library, and that still generates the same sequences in the same proportions. The same probability distribution over sequences is also achievable even if the state sequences end up oversegmented using a library in which pieces of the blocks are crossed over patterns. However, the original block structure can be retrieved by following the jumps where only one of the transition $\theta_k^{ij}$ is different than zero and connecting the pieces. Finally, the end goal *is* building the pdf model, and the model structure is simply used to constrain the estimation from a small dataset.

Transition probabilities different from zero or one:
$\theta_1 = [\frac{1}{4} \frac{3}{4} 0]$, $\theta_3^{12} = \theta_3^{13} = \frac{1}{2}$, $\theta_3^{22} = \frac{1}{3}$, $\theta_3^{21} = \frac{2}{3}$.

| Patterns | $R_1$ | A | C | G | C | G | G | A |
|---|---|---|---|---|---|---|---|---|
| | $R_2$ | C | A | A | C | G | C | A |
| | $R_3$ | A | C | A | G | A | C | T |
| Distribution over sequences | s | 1 | 1 | 1 | 2 | 2 | 2 | 2 |
| $\frac{1}{8}$ | x | A | C | G | C | G | C | A |
| | s | 2 | 2 | 2 | 1 | 1 | 1 | 1 |
| $\frac{1}{2}$ | x | C | A | A | C | G | G | A |
| | s | 1 | 1 | 1 | 3 | 3 | 3 | 3 |
| $\frac{1}{8}$ | x | A | C | G | G | A | C | T |
| | s | 2 | 2 | 2 | 2 | 2 | 2 | 2 |
| $\frac{1}{4}$ | x | C | A | A | C | G | C | A |

resented by forward and backward probabilities computed by the forward backward algorithm. At the same time, the forward and backward probabilities can be used to compute the likelihood $p(\{x_k\}_{k=1}^N|c)$. We will also be interested in marginals for each locus $p(s_k|\{x_k\}_{k=1}^N,c)$, as well as for the pairs of neighboring loci

$$p(s_k, s_{k-1}|\{x_k\}_{k=1}^N, c),$$

both of which can be computed from the forward and backward probabilities as well [17].

Having computed the posterior for both cases $c = 1$ and $c = 0$, the probability of disease can be computed as

$$p(c=1|\{x_k\}_{k=1}^N) = \frac{p(\{x_k\}_{k=1}^N|c=1)p(c=1)}{\sum_{v \in \{0,1\}} p(\{x_k\}_{k=1}^N|c=v)p(c=v)}.$$

(2)

The model parameters, including the pattern library $R_1, .., R_L$, and the transition parameters are computed using a learning algorithm that analyzes a set of sequences. In other words, for the example in the Table 1 the goal of learning would be to infer the pattern library from the generated sequences. In our experiments, the variability in the sequences is much larger, and yet they can still be captured well with only a handful of patterns. For the model described in this section, the learning can be performed using the EM algorithm, which uses the computed posteriors for all sequences to re-estimate the transition parameters $\theta$ in a standard HMM learning fashion, except that the statistics are collected for each site individually, as we assume non-homogeneous transitions in order to discover the block boundaries and use them to better constrain the distribution over the sequences:

$$\theta_{kc}^{mn} = \frac{\sum_{j=1}^J p(s_k = m, s_{k+1} = n|\{x_k^j\}_{k=1}^N, c)}{\sum_{j=1}^J p(s_k = m|\{x_k^j\}_{k=1}^N, c)}, \quad (3)$$

where $j$ enumerate the observed sequences (in our case for example, we had 387 people in the database).

In addition, we re-estimate the pattern libraries in the M step, as well, by collecting the statistics on which observed letter $x_k$ was associated with each pattern letter $R_k(s_k)$. Early in the learning, $x_k$ may not always correspond to $R_k(s_k)$ as we assume higher noise levels and the integration over the possible transition paths enforces an assignment to the pattern even when there are discrepancies. The following update rule sets each letter in the pattern library to the most frequently observed sequence letter $x_k$ associated with the pattern:

$$R_k^i = \underset{a \in \{A,C,G,T\}}{argmax} \sum_{j=1}^J [x_k^j = a] p(s_k^j = i|\{x_k^j\}_{k=1}^N), \quad (4)$$

where [] is an indicator function, i.e. [true]=1, [false]=0.

Note that the letter * can be added to the set {A, C, G, T} of letters in the library to model the missing values, or the missing value can be integrated over naturally in our framework, by simply treating the appropriate $x_k$ as hidden and performing an inference on it using the same forward backward procedure.

## 3 Modelling sequences of unordered letter pairs

The process used for observing the SNP structure cannot assign the observed letters to individual chromosomes in the chromosome pair. For instance, if the observed pair of nucleotides at a certain site is C,T, then each of the chromosomes could have had either one of the letters. The consistent ordering of the measured nucleotide pairs across the chromosome is not available to break the pairs into two nucleotide sequences, each adhering to the structure modelled in the previous section.

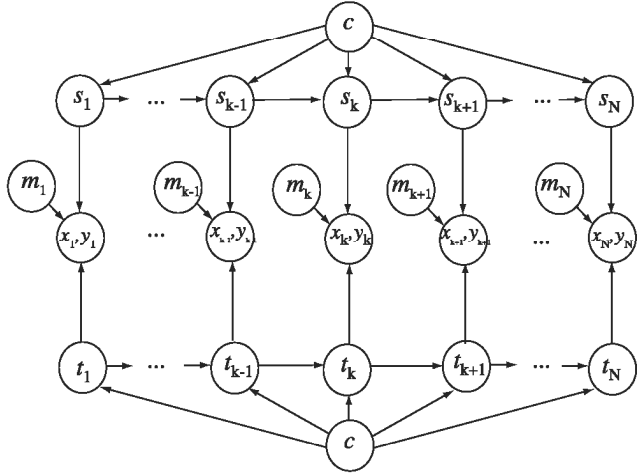

Figure 2: A generative model of unordered pairs of nucleotides, describing the data captured by a sequencing process in which the assignment of the nucleotides to individual chromosomes is not possible without additional information. By jointly estimating the block structure of the haplotypes and inferring the nucleotide assignment in a probabilistic framework, it is possible to use such data directly, without preprocessing, for model training and subsequent disease prediction.

However, if the pattern library is known, it is possible to infer the ordering in each pair (with some uncertainty) based on the expected patterns in the chromosomes. On the other hand, if the patterns are not known yet, than the observed pairs need to be ordered before the patterns can be learned. This can be resolved by iterating the two inference procedures, which can be formally defined as minimizing the variational free energy of a model that treats the ordering at each site $k$ as a hidden variable, $m_k$ (Fig. 2).

The two chains in the model are equivalent to the model in the previous section, while the observation likelihood for the pair of letters x,y is defined as,

$$p(x_k, y_k | s_k, t_k, m = 1) = p(x_k | s_k) p(y_k | t_k),$$
$$p(x_k, y_k | s_k, t_k, m = 0) = p(y_k | s_k) p(x_k | t_k), \quad (5)$$

where

$$p(z_k | s_k) = p(z_k | R_k^{s_k}), \quad p(z_k | t_k) = p(z_k | R_k^{t_k}), \quad (6)$$

for z=x or z=y, i.e., the pattern library is shared.

The prior on $m_k$ is assumed uniform, i.e., $p(m_k) = \frac{1}{2}$.

### 3.1 The free energy and variational inference and learning

The free energy of the model is defined as a function of the joint probability distribution over all variables and an auxiliary probability distribution defined only over the hidden variables $Q(\{m_k, s_k, t_k\}_{k=1}^N, c)$:

$$F = -\sum Q(\{m_k, s_k, t_k\}, c) \log \frac{Q(\{m_k, s_k, t_k\}, c)}{P(\{x_k, m_k, s_k, t_k\}, c)}.$$

The free energy is limited from below by the negative log likelihood of the data $-\log p(\{x_k\})$, and the bound becomes tight when Q is equal to the true posterior over the hidden variables[16]. The variational learning consists of successive minimizations of the free energy with respect to a constrained posterior Q and the model parameters $\theta, R$.

Constrained forms of the function Q are used to make the inference tractable. For our model, the number of possible configurations of $\{m_k\}$ is exponential, so we use the following form for Q:

$$Q = Q(\{s_k\}_{k=1}^N) Q(\{t_k\}_{k=1}^N) \prod_k Q(m_k). \quad (8)$$

Using this form, the optimization breaks into iteratively solving for $Q(m_k)$ and running the forward-backward based inference and learning in the chains with the equivalent observation log likelihoods

$$\log \tilde{p}(x_k, y_k | s_k) = Q(m_k = 1) \log p(x_k | s_k) +$$
$$Q(m_k = 0) \log p(y_k | s_k)$$
$$\log \tilde{p}(x_k, y_k | t_k) = Q(m_k = 1) \log p(y_k | s_k) +$$
$$Q(m_k = 0) \log p(x_k | t_k).$$

The forward-backward algorithm provides the marginals of the posteriors $Q(\{s_k\}) = \tilde{p}(\{s_k\} | \{x_k, y_k\})$ and $Q(\{t_k\}) = \tilde{p}(\{t_k\} | \{x_k, y_k\})$, based on the above equivalent observation likelihoods and the current estimate of the pattern library and transition probabilities.

The update rule for the distribution over assignment variables is:

$$Q(m_k = 1) \propto exp \left( \sum_{s_k} Q(s_k) \log p(x_k | s_k) + \sum_{t_k} Q(t_k) \log p(y_k | t_k) \right)$$

$$Q(m_k = 0) \propto exp \left( \sum_{s_k} Q(s_k) \log p(y_k | s_k) + \sum_{t_k} Q(t_k) \log p(x_k | t_k) \right),$$

where $Q(m_k)$ is normalized so that $Q(m_k = 1) + Q(m_k = 0) = 1$, and the marginals $Q(s_k)$ and $Q(t_k)$ are computed using the forward-backward procedure.

In inference, these sets of updates are iterated until convergence, starting with the uniform $Q(m_k = 1) = 0.5$. Model parameters can be updated after each update of the forward and backward distributions, leading to an efficient learning algorithm.

## 4 Training regimes and local maxima

The described models have a large number of parameters but a constrained structure. The following set of principles should be followed in order to avoid local maxima:

- Initially, the noise in the observation likelihood should be set to a large value in order to avoid getting peaked posteriors early in the learning. Peaky distributions can quickly lead to overtraining.

- The number of patterns in the library is defined beforehand and should be kept 3-4 patterns bigger than the maximum expected number of haplotype blocks at any site. Again, the reason is to avoid local maxima.

- The transition probabilities should be initialized to strongly favor staying in the same class everywhere, with the distribution over the possible places to escape uniform.

- The pattern library should be shared between the two index (allele) sequences $s$ and $t$, and appropriately updated based on the inference in both chains.

## 5 Results

We evaluated our model's performance on a classification task of detecting susceptibility to Crohn's disease from previously unseen SNP sequences. In order to further assess quality of our model we compared it to a Naive Bayes classifier, which models distribution of alleles on each SNP locus independently. The comparison was conducted in 5 different experiments, for each of which a test set and training set of sizes 100 and 287 were randomly selected from the pool of individuals previously used by [2]. The test sets consisted of SNP sequences for 50 healthy and 50 affected individuals. Both the training and the test set were presented to the algorithm without their parental sequences. Both classifiers were trained on the same training set and then evaluated on the same test set. The error rate was computed by adding false positives and false negatives rates, which were roughly equal. The error rates for classifier based on our model had a mean of 20.4% and 1.5% standard deviation, while Naive Bayes classifier had a mean of 40.6% and a standard deviation of 0.9%. We have performed a paired t-test on these data and obtained p-value of %5.7159e-009. This indicates that the inferred block structure indeed was a good basis for modelling the variability in the data. Visual inspection revealed that our segments agreed with the ones shown in [2] in about 50% of the hotspots, while the differences were mostly due to over parameterization with 7 patterns, leading to learning multiple copies of the same block (Fig. 3). While this structure captures the variability in the data as it is, it can be further cleaned up by following the transitions in the chain and linking the pieces together as discussed above, in order to compare with the algorithms that had access to ancestral states. We compare the block structure we inferred in the dataset without the parental information with the block structure inferred from the pedigree data[2] at www.research.microsoft.com/~jojic/haplo.html.

## 6 Conclusions

By modelling various causes of variability jointly, we were able to develop learning algorithms that simultaneously extract the block structure of the haplotypes and learn to predict the Chrohn's disease from new SNP sequences. The model can automatically deal with the fact that the observed nucleotide pairs are unordered, as well with the missing data. The model structure is used as a guide for building the model, but the main goal was achieving a good pdf estimate to be used in classification when only a small training set is available (in comparison to the $4^{103}$ possible 103-long sequences). Thus it was not crucial to estimate the block structure so as to optimize an MDL criterion.

Even though both the training and testing were performed ignoring the parent data, the model was able to infer the common block structure in the data. This is one of the major advantages of graphical models over generic classification algorithms, such as neural nets - the structure of the generative model mimics the known, or accepted structure of the data at hand, thus making it easier to interpret the results beyond the classification rate.

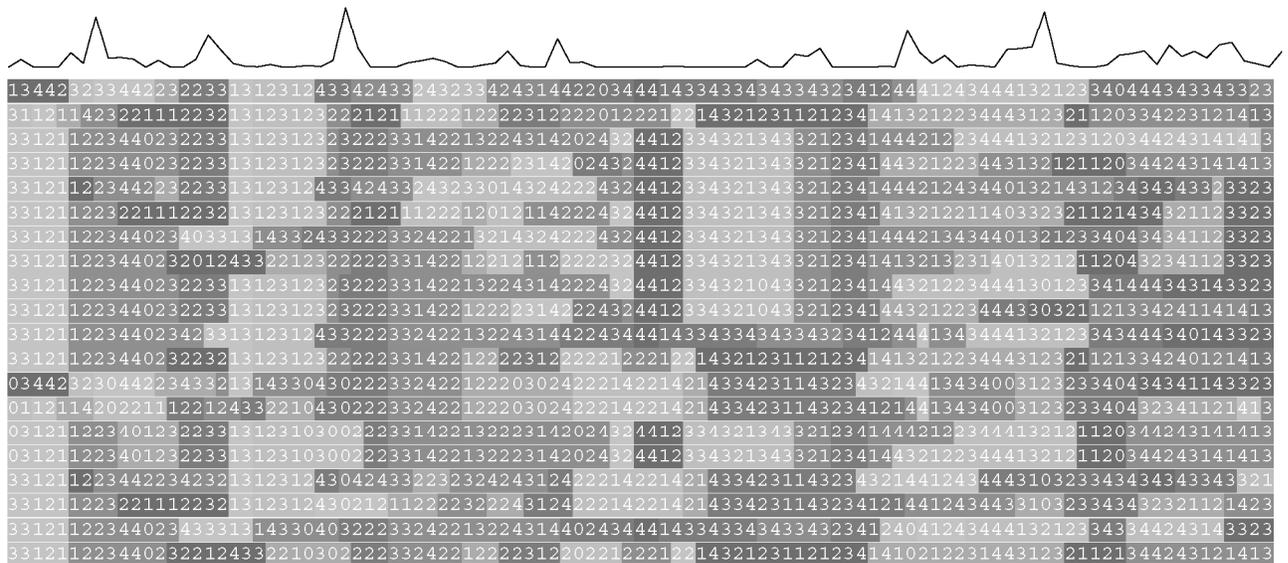

A fraction of the training sequences and the mapping to the pattern library

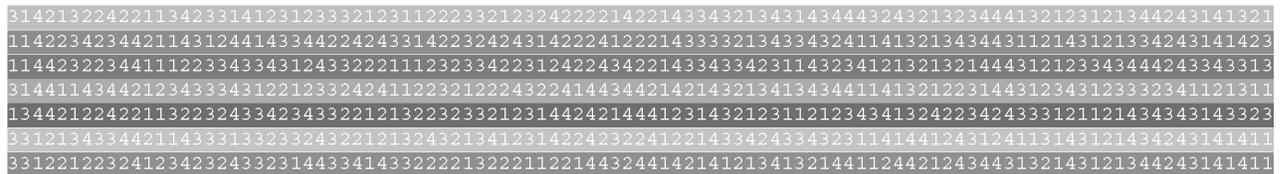

The learned patterns

Figure 3: A few sequences from the training set, and the learned patterns. In combination with the transition probabilities, these patterns represent the probability distribution function over the data even without compacting the blocks into a more regular structure. On top, we also show the entropy of the transition distribution for each site. High entropy indicates likely branching and could be used as an initial guess for recombination hotspots, although better picture can be obtained by the block concatenation procedure described in the text.